\documentclass{SCIS2025}
\begin{document}
\ArticleType{PERSPECTIVE}
\Year{2024}
\Month{}
\Vol{}
\No{}
\DOI{}
\ArtNo{}
\ReceiveDate{}
\ReviseDate{}
\AcceptDate{}
\OnlineDate{}
\AuthorMark{}
\AuthorCitation{}
\title{An Infrastructure Software Perspective Toward Computation Offloading between Executable Specifications and Foundation Models}{Computation Offloading between Executable Specifications and Foundation Models}

\author[1]{Dezhi Ran}{}
\author[2]{Mengzhou Wu}{}
\author[2]{Yuan Cao}{}
\author[3]{Assaf Marron}{}
\author[3]{David Harel}{}
\author[1]{Tao Xie}{{taoxie@pku.edu.cn}}

\address[1]{Key Lab of HCST (PKU), MOE; SCS, Peking University, Beijing 100871, China}
\address[2]{School of EECS, Peking University, Beijing 100871, China}
\address[3]{Dept. of Computer Science and Applied Mathematics, Weizmann Institute of Science, Rehovot 6 - 7632706, Israel}

\maketitle

\begin{multicols}{2}
\noindent 
Foundation Models (in short as \textit{FMs})~\cite{1} have revolutionized software development and become the core components of large software systems.
This paradigm shift, however, demands fundamental re-imagining of software engineering theories and methodologies~\cite{3}. 
Instead of replacing existing software modules implemented by symbolic logic, incorporating FMs' capabilities to build software systems requires entirely new modules that leverage the unique capabilities of FMs.
Specifically, while FMs excel at handling uncertainty, recognizing patterns, and processing unstructured data, we need new engineering theories that support the paradigm shift from explicitly programming and maintaining user-defined symbolic logic to creating rich, expressive requirements that FMs can accurately perceive and implement.

In this article, we present a new perspective for building reliable FM-based software systems via \textit{computation offloading}, which strategically distributes computational tasks between FMs and traditional executable specifications~\cite{4} (e.g., symbolic programs) based on their respective strengths~\cite{5,6}. 
Computation offloading has been widely adopted by cloud computing~\cite{7}, which primarily focuses on distributing computational tasks based on hardware resources and performance metrics across cloud and edge devices.
We extend this concept to a semantic level, where offloading decisions are primarily driven by the inherent capabilities of FMs and executable specifications rather than just computational capacity.
While FMs excel at pattern recognition and handling unstructured data, they may struggle with precise logical reasoning and strict-constraint satisfaction. Conversely, traditional software components are adept at exact computations and formal verification but lack the flexibility and learning capabilities of FMs. 
By intelligently offloading computations between these two complementary paradigms, we can create a robust and efficient software system that leverages the best of both worlds. This offloading mechanism serves as a fundamental building block for FM-based software, enabling developers to focus on high-level specifications while the system automatically determines the optimal execution strategy across FMs and traditional components.

\begin{figure*}[t]
\centering
\includegraphics[width=0.95\linewidth]{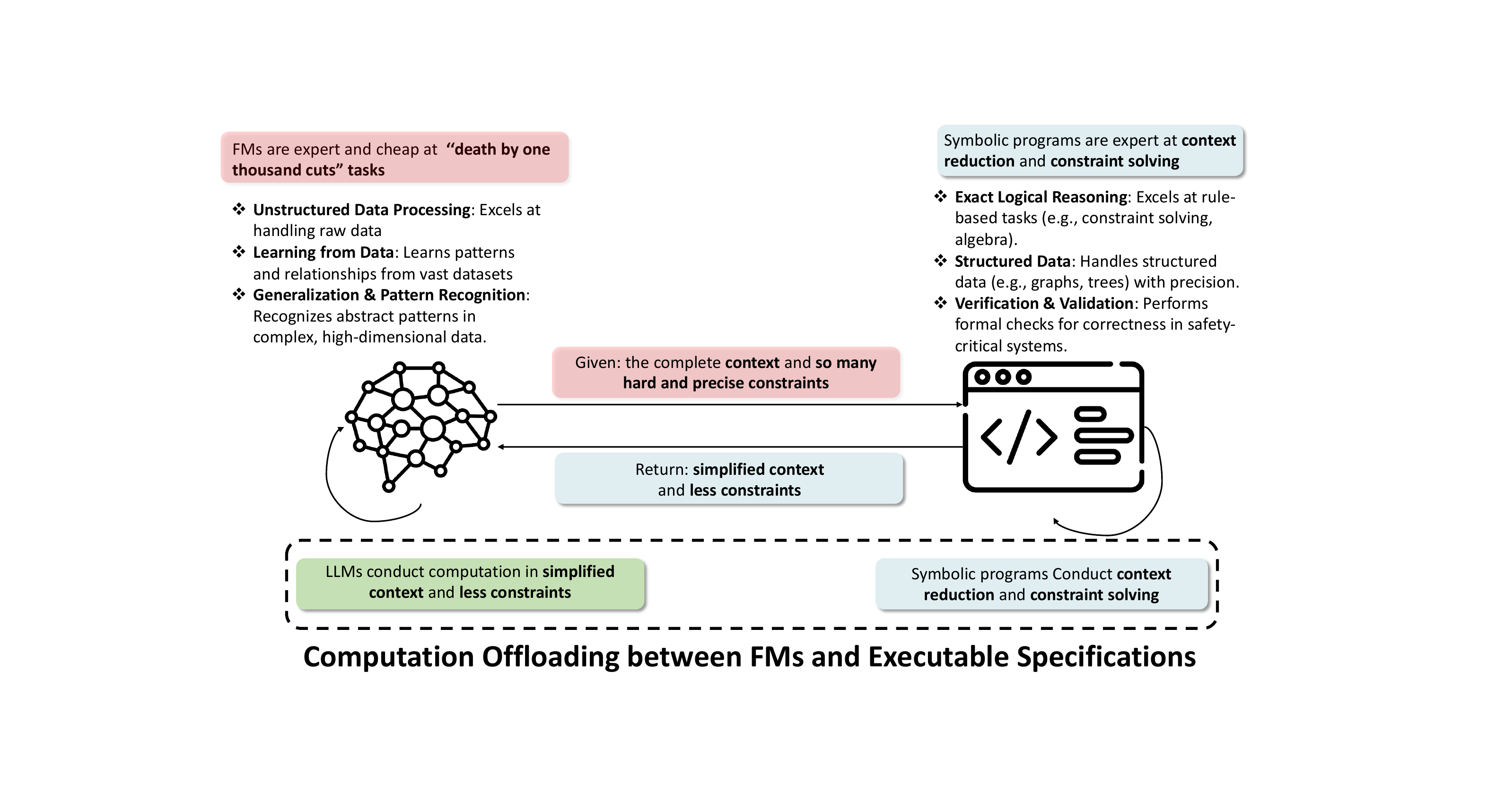}
\caption{Computation Offloading between Executable Specifications (Symbolic Programs) and Foundation Models.}
\label{fig1}
\end{figure*}

\lettersection{Complementarity between FMs and Executable Specifications}
FMs and executable specifications such as symbolic programs handle different types of computation tasks and actually complement with each other. 
For example, in image processing, FMs can identify complex patterns while symbolic programs can manipulate precise geometric transformations. 
In natural language processing, FMs can generate creative and coherent texts while symbolic programs can conduct rigorous grammar checking. 
This symbiotic relationship enables a new paradigm of software development where each approach's strengths compensate for the other's weaknesses.

\noindent\textbf{Pros and cons of FMs.}
From a usage convenience perspective, FMs excel at handling unstructured data and large-scale pattern recognition tasks, making them particularly effective at managing varied inputs and corner cases that traditionally would require numerous individual handling rules. Their learning curves are relatively gentle for basic usage, as they can understand natural language instructions and generate responses without requiring extensive programming knowledge. Development costs are initially high due to the need for substantial computational resources and training data, but they become cost-effective when handling a large number of similar tasks. Maintenance costs can be significant, particularly for model updates and retraining. The main maintenance challenge of FMs lies in their expandability - while FMs can be fine-tuned for new tasks, ensuring consistent performance across different domains and maintaining model quality require careful management of training data and model architecture.

\noindent\textbf{Pros and cons of executable specifications.}
Executable specifications, manifested as symbolic programs, offer precise control and verifiable behavior but come with their own trade-offs. 
Their usage convenience is highest when dealing with well-defined, structured problems and formal specifications. The learning curve is steep, requiring deep understanding of programming languages, formal methods, and domain-specific knowledge. Development costs are moderate for simple specifications but can escalate rapidly for complex systems requiring formal verification. Maintenance costs of executable specifications are generally lower than those of FMs, as changes can be implemented through direct code modifications. The  expandability of executable specifications is excellent for structured problems within their domain, allowing for modular development and clear separation of concerns. However, they struggle with ambiguous or poorly defined requirements and lack the natural flexibility of FMs in handling variations in input data.

\lettersection{Orchestrating Computation Offloading via Infrastructure Software}
To enable computation offloading, we envision infrastructure software~\cite{8, 9}, a sophisticated middleware layer that serves as the foundation for developing and maintaining FM-native software. 
The infrastructure software orchestrates computation offloading through a sophisticated interplay of multiple components, each designed to optimize the allocation and execution of tasks across FMs and symbolic programs. This orchestration process involves the following eight key mechanisms.

\noindent\textbf{Context analysis and task decomposition.}
The infrastructure software first analyzes the computational context and requirements of incoming tasks. This analysis involves identifying the nature of the computation, such as whether it requires pattern recognition, logical reasoning, or constraint solving. Complex tasks are decomposed into subtasks that can be efficiently handled by either FMs or symbolic programs. The analysis also considers data characteristics and constraints that influence the offloading decision, ensuring optimal task distribution.

\noindent\textbf{Resource assessment and allocation.}
Based on the context analysis, the infrastructure software performs dynamic resource assessment. This assessment includes evaluating the availability and capabilities of different FMs, analyzing the computational resources required for symbolic-program execution, and considering performance requirements, latency constraints, and resource costs. The system then determines optimal strategies of resource allocation  for different subtasks, ensuring efficient resource utilization across the entire system.

\noindent\textbf{Intelligent offloading decisions.}
The infrastructure software makes informed decisions about where to execute each computation. Tasks of pattern recognition and unstructured data processing are directed to appropriate FMs, while tasks of precise logical operations and formal verification  are assigned to symbolic programs. Hybrid tasks may be split across both paradigms with careful coordination, leveraging the strengths of each approach while minimizing their limitations.

\noindent\textbf{Data flow management.}
Efficient data handling is crucial for successful computation offloading. The infrastructure software manages data transformation between FM-compatible and symbolic program formats, implements efficient data transfer protocols between components, and maintains data consistency across different computational paradigms. 
In addition, the data integrity and privacy during the data transfer and transformation should be secured.

\noindent\textbf{Execution coordination.}
The infrastructure software coordinates the execution of offloaded computations through sophisticated scheduling and synchronization mechanisms. This coordination includes managing the parallel execution of independent subtasks, handling inter-component communication and state management, and implementing robust failure recovery and error-handling mechanisms. The coordination ensures smooth operation across all system components.

\noindent\textbf{Quality assurance and verification.}
Throughout the execution process, the infrastructure software ensures quality and correctness through continuous monitoring of execution quality and performance metrics. It verifies results against specified requirements and constraints, implements runtime validation of FM outputs, and ensures compliance with security and privacy requirements. This comprehensive quality control ensures reliable system operation.

\noindent\textbf{Adaptive optimization.}
The infrastructure software continuously learns and adapts its orchestration strategies based on operational experience. It learns from previous execution patterns to improve offloading decisions, adapts to changing resource availability and system conditions, and optimizes performance based on observed metrics and feedback. This adaptive mechanism  ensures increasingly efficient operation over time.

\noindent\textbf{Developer interfaces.}
The infrastructure software provides intuitive interfaces for developers through high-level abstractions for specifying computational requirements. It offers tools for monitoring and controlling the offloading process, APIs for customizing offloading strategies, and comprehensive debugging and profiling capabilities for hybrid computations.

Through these eight key mechanisms, the infrastructure software creates a seamless environment where computation offloading becomes transparent to developers while maintaining optimal performance and reliability. This orchestration layer abstracts away the complexity of managing hybrid computations, allowing developers to focus on their primary task of specifying desired behaviors and requirements. The result is an  efficient and accessible software development process that leverages the strengths of both FMs and symbolic programs while minimizing their respective limitations.

\lettersection{Technical Challenges}
Realizing the vision of infrastructure software that bridges FMs  and executable specifications needs to address the following six  fundamental technical challenges, spanning multiple layers of the system stack and requiring innovative solutions that combine insights from machine learning, software engineering, and systems research.

\noindent\textbf{Semantic-gap resolution.}
One of the most fundamental challenges lies in bridging the semantic gap between FMs and executable specifications. A key issue is the representation mismatch, as FMs operate on probabilistic, distributed representations while executable specifications require precise, deterministic inputs. Developing reliable translation mechanisms between these paradigms remains challenging. Context preservation presents another significant hurdle, requiring careful balance to ensure that essential context is preserved when simplifying problems for FMs while maintaining the precision required by executable specifications. Additionally, maintaining semantic consistency across different levels of abstraction and between different components of the system poses further challenges.

\noindent\textbf{Resource management and optimization.}
The heterogeneous nature of computation introduces challenges of complex resource management. Dynamic resource allocation becomes critical for efficiently distributing computational resources between FM inference and symbolic execution based on changing workloads and requirements. Memory management must handle the diverse memory access patterns and requirements of both FMs and executable specifications while maintaining system performance. Energy efficiency considerations also come into play when optimizing the energy consumption of hybrid systems that combine both FM inference and traditional computation.

\noindent\textbf{Reliability and verification.}
Ensuring system reliability presents unique challenges for such hybrid systems. 
Developing solutions to verify the correctness of computations that combine probabilistic FM outputs with deterministic executable specifications remains a significant challenge. 
Understanding and controlling error propagation between FM outputs and executable specifications require careful consideration.
Efficient mechanisms of runtime verification that can handle both FM-based and specification-based components adds another layer of complexity to the system design.

\noindent\textbf{Performance and scalability.}
The scalability and complexity of modern software systems introduce significant performance challenges. It is crucial to minimize the latency overhead of computation offloading and context switching between FMs and executable specifications. The infrastructure software must scale effectively with increasing system complexity and data volume. This scaling requires developing efficient caching and prediction mechanisms to reduce unnecessary computation and data transfer while maintaining system responsiveness.

\noindent\textbf{Security and privacy.}
The integration of FMs with executable specifications within a system raises new security concerns. Data protection becomes more complex as sensitive information moves between different components of the system. The increased attack surface that comes from combining multiple computational paradigms requires careful management. Privacy preservation mechanisms must be developed to maintain confidentiality when processing sensitive data through FMs.

\noindent\textbf{Development and deployment.}
The practical aspects of building and deploying hybrid FM-specification   systems face multiple  challenges. Creating effective tools and environments that support the development of such  systems requires new techniques to software development. For example,  testing and debugging techniques must evolve to handle systems that combine deterministic and probabilistic components. Version control and management systems need to adapt to handle the coordinated evolution of both FMs and executable specifications.

\lettersection{Research Directions}
Tackling the preceding challenges calls for multiple critical research directions for future research efforts. The development of formal methods capable of reasoning about hybrid FM-specification systems represents a crucial area of investigation. New programming models that naturally support computation offloading need to be created. Automated optimization techniques for resource allocation and performance tuning require further research. Verification frameworks that can handle both probabilistic and deterministic components must be designed. Research into human-AI interaction patterns for software development in this new paradigm is also  essential for realizing the full potential of these systems.

\Acknowledgements{This work was supported by National Natural Science Foundation of China (NSFC) and Israel Science
Foundation (ISF) (Grant No. 62161146003 and 3698/21). Additional support was by National Natural Science Foundation of China (Grant No. 623B2006 and Grant No. 92464301).}

\end{multicols}
\end{document}